\documentclass[aps,twocolumn,showpacs,superscriptaddress,citeautoscript,prl,floatfix]{revtex4-2}

\usepackage{graphicx} %
\usepackage{amsmath}
\usepackage{amsfonts}
\usepackage{tabularx}
\usepackage{cancel}
\usepackage{bbold} 
\usepackage{wrapfig}
\usepackage{amssymb}
\usepackage{bm}
\usepackage[T1]{fontenc}
\usepackage{textcomp}
\usepackage{enumitem}
\usepackage{cancel}
\usepackage{bbold}
\usepackage{esvect}
\usepackage{commath}
\usepackage[most]{tcolorbox}
\usepackage{lipsum, babel}
\usepackage{comment}
\usepackage{lineno}
\usepackage{xcolor}

\usepackage[breaklinks, hidelinks]{hyperref}
\hypersetup{
    colorlinks,
    linkcolor={red!80!black},
    citecolor={red!80!black},
    urlcolor={red!80!black}
}

\usepackage{adjustbox}
\graphicspath{{figures/}}

\begin{document}

\title{Many-body quantum chaos in excitonic spectra from first principles}

\author{Daniel Hernang\'omez-P\'erez}
\email{d.hernangomez@nanogune.eu}

\affiliation{%
CIC nanoGUNE BRTA, Tolosa Hiribidea 76, 20018 San Sebastián, Spain
}%

\author{Rafael A. Molina}
\email{rafael.molina@csic.es}

\affiliation{Instituto de Estructura de la Materia IEM-CSIC, Serrano 123, E-28006 Madrid, Spain}

\date{\today}

\begin{abstract}
    We demonstrate that realistic excitonic many-body Hamiltonians obtained from first-principles GW-Bethe–Salpeter equation calculations can exhibit quantum chaos governed by random-matrix universality.
    Considering a prototypical van der Waals heterostructure (WS\textsubscript{2}-graphene), with and without lattice disorder, we analyze their energy-resolved spectral correlations %
    and identify a disorder-driven crossover from regular to complete chaotic dynamics. %
    We show that while pristine samples exhibit incomplete chaos (non-ergodicity) due to an approximate valley symmetry that restricts excitonic mixing, the presence of disorder-induced electronic flat bands act as a catalyst for valley mixing to drive the system into a fully developed chaotic (ergodic) regime with reduced symmetry.
    Crucially, fluctuations in many-body oscillator strengths are shown to follow universal Porter–Thomas statistics, directly linking the underlying quantum chaos and experimentally accessible optical observables. 
    Finally, by examining long-range spectral correlations, we estimate the Thouless time associated to excitonic mixing across the entire many-body bandwidth.
    Our results establish excitons as a highly tunable platform for probing many-body ergodicity and its spectroscopic signatures in realistic interacting 2D materials. 
\end{abstract}

\maketitle

\subparagraph{\em Introduction.} 
Quantum chaos investigates universal spectral and dynamical signatures of non-integrability in quantum systems. 
An important result for such systems, whose classical counterparts are chaotic, is that eigenvalue fluctuations can be described by random-matrix theory (RMT), whereas integrable systems follow Poisson statistics \cite{BerryTabor1977,Bohigas1984,Haake_book}.
In many-body systems, chaos plays a central role in thermalization, entanglement dynamics, and information scrambling, with RMT and the eigenstate thermalization hypothesis providing key diagnostics \cite{Deutsch1991,Srednicki1994,Gomez2011,DAlessio2016}. 
Yet, deviations from universality are in principle relatively common, arising as a consequence of locality, conservation laws, or system geometry. Prominent examples include constrained systems with quantum scars \cite{Serbyn2016, Moudgalya2022}, long-range interacting model systems \cite{Xu2019,Else2020}, and systems near many-body localization \cite{Nandkishore2015,Serbyn2016,Corps2020,Sierant2025}.

From an experimental standpoint, signatures of quantum chaos in solid-state systems have been  restricted to a small number of highly controlled, few-body platforms. An example is the Rydberg exciton series in Cu$_2$O, where the combination of band anisotropy, spin-orbit coupling and external  magnetic fields yield a non-integrable two-particle Hamiltonian with Wigner-Dyson level statistics and GOE (Gaussian Orthogonal Ensemble) to GUE (Gaussian Unitary Ensemble) crossovers \cite{Assmann2016,Schweiner2017}.
Additional realizations include mesoscopic systems and quantum dots, whose transport properties exhibit universal fluctuations \cite{Jalabert1990,Marcus1992,Alhassid2000}, or microwave billiards emulating chaotic quantum Hamiltonians with high fidelity \cite{Dietz2015}.
However, these systems are effectively described by single-particle or few-body dynamics, for which RMT universality is well established. Direct evidence of true many-body quantum chaos in crystalline solid-state materials remains elusive due the difficulty of accessing their many-body spectral correlations. 

Two-dimensional (2D) atomically thin van der Waals (vdW) materials \cite{Wang2012, Geim2013, CastroNeto2016} provide an attractive setting for exploring many-body quantum chaos. In these systems, dimensionality and nonlocal dielectric screening produce remarkably strong Coulomb interactions, giving rise to strongly bound excitons \cite{Mak2010, Qiu2013, Chernikov2014, Ugeda2014, Wang2018} with binding energy orders of magnitude larger than in bulk semiconductors. 
These excitons also possess a complex spectral structure resulting from symmetry-rich Hilbert spaces, highly sensitive to disorder, interlayer coupling or other external perturbation \cite{Echeverry2016, Robert2017, Molas2017, Wang2018, Molas2019, Kundu2023, Richter2024}.
Theoretically, excitonic spectra can be described via first-principles GW-Bethe–Salpeter (GW-BSE) approaches \cite{Hedin1965, Hybertsen1985, Hybertsen1986, Albrecht1998, Rohlfing1998, Rohlfing2000, Deslippe2012}. %
Crucially, while the BSE is formally cast as an eigenvalue equation in a two-particle electron-hole basis \cite{Rohlfing2000, Deslippe2012}, the interaction kernel inherits the full non-local, dynamic screening of the solid. The resulting excitonic fine structure represents an effective many-body configuration space rather than a simple isolated two-body problem, with Hamiltonian matrix dimensions \cite{Hernangomez2023} comparable to those used to study quantum chaos in nuclear physics or many-body lattice models. Yet, the statistical properties of these realistic \textit{ab initio} Hamiltonians have remained largely unexplored.
%

In this Letter, we demonstrate the emergence of many-body quantum chaos in the excitonic spectra of 2D vdW heterostructures. Using first-principles GW-BSE calculations, we show that the complex electronic screening embedded in the \textit{ab initio} excitonic Hamiltonian generates spectral correlations governed by RMT. 
We find that pristine heterostructures exhibit incomplete chaos due to an approximate valley symmetry that suppresses full excitonic mixing, preventing full ergodicity and leading to non-ergodic spectral statistics. In contrast, vacancy-induced flat bands enhance valley hybridization, lifting this symmetry and driving the system into a fully ergodic regime. 
We further demonstrate that the fluctuations of the many-body oscillator strengths follow universal Porter–Thomas statistics and extract their energy-dependent characteristic exponent. Analysis of long-range spectral correlations additionally reveals the Thouless energy associated with excitonic valley mixing across the many-body bandwidth.
Our results establish \textit{ab initio} many-body spectra as a realistic platform for probing many-body ergodicity and its breakdown, as well as providing a direct link between these phenomena and experimentally accessible optical observables.

\subparagraph{\em Model and methods.} We consider %
a prototypical 2D vdW heterobilayer (WS$_2$-graphene heterobilayer), with and without monoatomic chalcogen vacancies. The sulfur vacancy models structural or lattice disorder. Their optical properties  were previously investigated in Ref. \cite{Hernangomez2023}, which provides comprehensive details of the \textit{ab initio} methodology. Here, we briefly summarize the essential elements relevant to the present statistical analysis.
Excitonic energies and many-body states, as well as the associated many-body oscillator or dipolar strengths, were obtained from first principles GW-BSE calculations \cite{Rohlfing1998, Rohlfing2000, Salpeter1951} performed with BerkeleyGW \cite{Deslippe2012}.
To characterize (many-body) quantum chaos, we analyze spectral correlations of the excitonic eigenenergies using spacing ratios and its associated statistics \cite{Oganesyan2007, Atas2013}. This approach avoids spectral unfolding while retaining the universal signatures of RMT. %
Fluctuations of the oscillator strengths, and associated eigenstates, were analyzed using Porter-Thomas statistics \cite{Porter1956}. Finally, long-range spectral correlations (higher-order) \cite{Corps2021} were examined employing the ensemble averaged power spectrum of cumulative level fluctuations, and the result used to estimate the Thouless energy \cite{Thouless1972}.
Additional technical details are provided in the End Matter and the Supporting Information (SI).

\begin{figure}
\centering
\includegraphics[width=1.0\columnwidth]{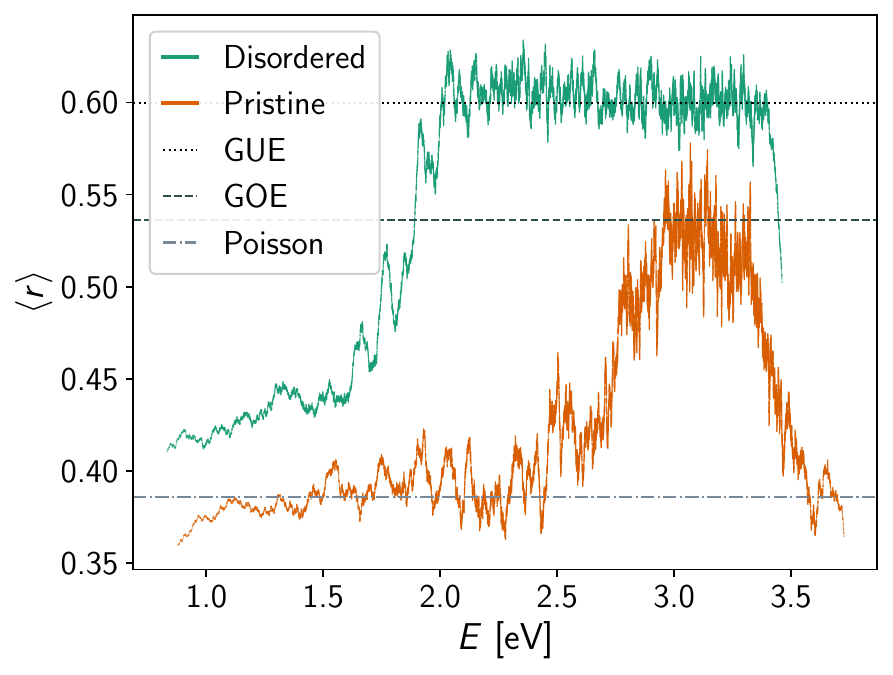}
\caption{Average adjacent level spacing ratio, $\langle r\rangle$, taken over windows of 1000 excitonic levels for the pristine (orange) and lattice disordered (green) WS$_2$-graphene heterobilayer. 
Black dotted, dark grey dashed, and light grey dash-dotted lines denote the theoretical limits for the GUE, GOE, and Poissonian statistics, respectively. 
}\label{fig:ratio_total}
\end{figure}

\subparagraph{\em Results.} 
Fig.~\ref{fig:ratio_total} shows the energy-resolved average
spacing ratio, $\langle r \rangle$, for the excitonic spectrum of the WS$_2$/graphene vdW heterostructure in the pristine (orange dots) and disordered (green dots) cases.
At low excitonic energies, we find that both systems exhibit values consistent with the Poisson limit ($\langle r \rangle \simeq 0.386$), indicating weak spectral correlations associated with exciton states relatively well localized.
As the exciton energy increases, $\langle r \rangle$ grows continuously, pointing towards an onset of level repulsion and subsequent emergence of quantum-chaotic behavior.
The crossover towards chaos, characterized by level repulsion and spectral rigidity, occurs substantially earlier in the lattice-disordered system, starting around $E \simeq 1.5$ eV, whereas the pristine heterostructure transitions at $E \simeq 2.5$ eV.
A more profound distinction emerges at higher exciton energies. The disordered many-body spectrum develops a remarkably robust, extended plateau that saturates around the GUE prediction ($\langle r \rangle \simeq 0.603$), while the pristine system only approaches transiently the GOE expected value ($\langle r \rangle \simeq 0.53$) and never exhibits a comparably extended chaotic window.
We note that this disorder induced GOE-to-GUE crossover highlights a subtle interplay between spin-orbit coupling (SOC) and symmetries. The ensemble difference suggests that the pristine heterostructure remains partially constrained by an effective antiunitary symmetry that balances the phases of the complex matrix elements, thus enforcing GOE statistics despite having explicit inclusion of the SOC in the calculations.
Previous works have identified an approximate valley symmetry \cite{Kleiner2024}, which constraints and limits excitonic mixing, preventing the development of fully universal spectral correlations (``incomplete'' chaos or weakly non-ergodic regime). %
Sulfur vacancies fundamentally alter this picture. Rather than only introducing simple single-particle intervalley scattering processes, the vacancy-induced electronic flat bands also compress the configuration energy scales and dramatically increase the density of multi-configurational electron-hole states interacting through the Coulomb potential \cite{Hernangomez2023}.
This induces strong mixing between excitonic configurations associated with different K-valleys,  creating a stronger level repulsion and allowing for the emergence of a robust fully-developed GUE plateau. In this sense, lattice disorder acts as a catalyst for many-body chaos, lifting the constraints on the valley symmetry by acting as an effective gauge field and driving the system from an incomplete chaotic regime towards a fully developed ergodic one governed by GUE statistics.
\begin{figure}
\centering
\includegraphics[width=0.95\columnwidth]{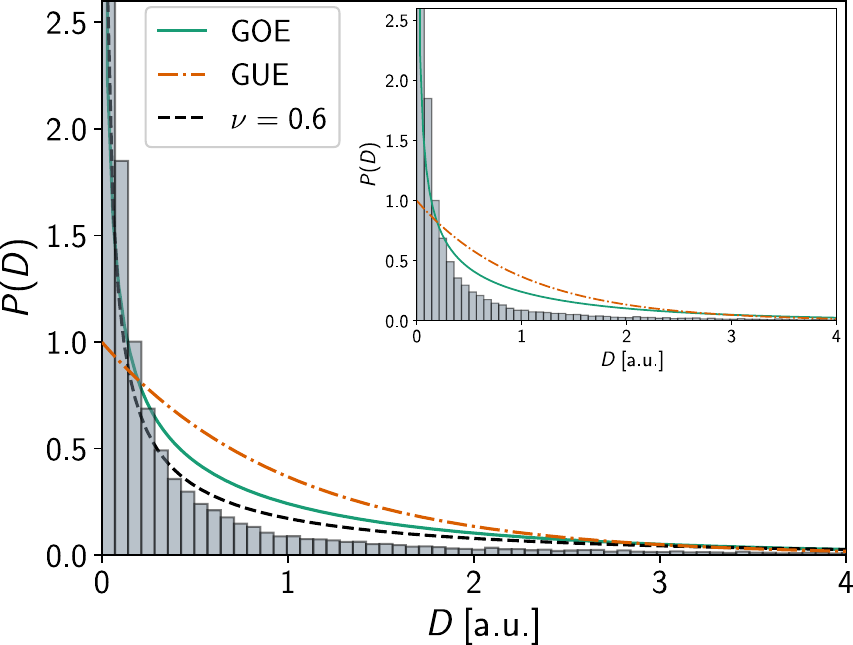}
\caption{Distribution of the normalized dipolar strength, $D/\left<D\right>$, for the pristine system. We compare our results with a $\chi^2$ distribution with $\nu$ degrees of freedom (dashed black line). The  solid green line corresponds to the Porter-Thomas distribution for the GOE  while the GUE result is shown by the dashed dotted orange line. The main panel shows data in the energy window $E \in [3.15,3.3]$ eV. The inset corresponds to the energy interval $E \in [1.5,2.0]$ eV. 
}
\label{fig:PT_clean}
\end{figure}

%
While spacing ratios probe short-range correlations between exciton energies, they do not directly reveal information on the ergodicity of the corresponding many-body quantum states.
In RMT, the distribution of such (normalized) random eigenstates also lead to universal statistics for transition amplitudes associated to optical intensities. These are described by generalized Porter-Thomas  distribution \cite{Porter1956}, which takes the form of a $\chi^2$ distribution with $\nu$ degrees of freedom,
\begin{equation}\label{Eq:PTnu}
P_\nu(x)
= \dfrac{\nu}{2x_0  \Gamma(\nu/2)}
  \left(\dfrac{\nu x}{2x_0}\right)^{\!\nu/2 - 1}
  \exp\!\left(-\dfrac{\nu x}{2x_0}\right),
\end{equation}
where $x_0$ is the mean value of the intensities and $\Gamma(x)$ is the Gamma function. In open systems, $\nu$ denotes the number of statistically independent channels that contribute to a decay process.
For closed systems, $\nu$ reflects underlying symmetries of the Hamiltonian, $\nu =1$ corresponds to the GOE and $\nu =2$ to the GUE (see further information in the End Matter).
\begin{figure}
\centering
\includegraphics[width=0.95\columnwidth]{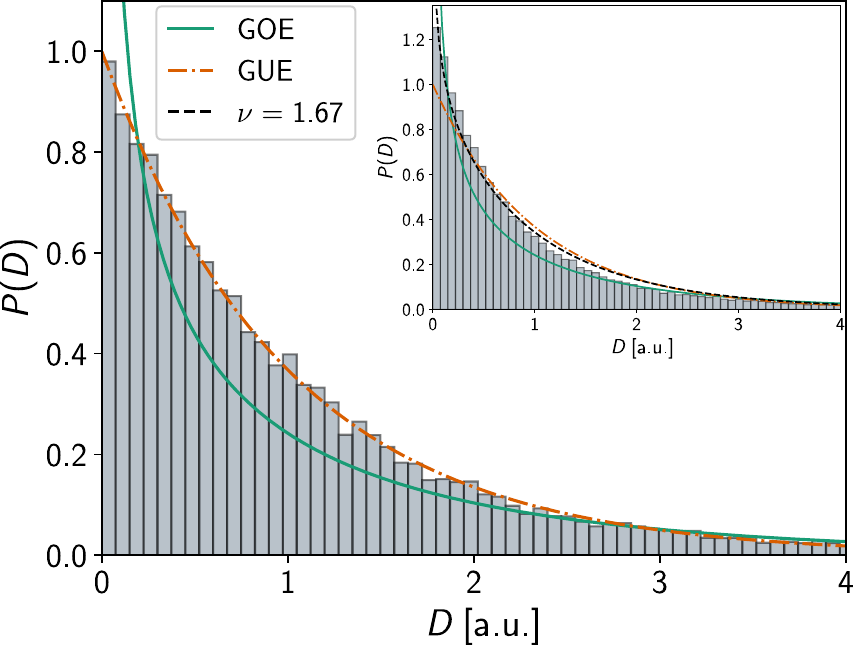}
\caption{Distribution of the normalized dipolar strength, $D/\left<D\right>$, for the lattice disordered system. We compare our results with a $\chi^2$ distribution with $\nu$ degrees of freedom (dashed black line), the dashed dotted orange line corresponds to the Porter-Thomas distribution for the GUE while solid green line corresponds to the GOE result. The main panel shows the distribution in the energy window $E \in [3.15,3.28]$ eV, with fully developed quantum chaos. The inset corresponds to the energy interval $E \in [2.49,2.68]$ eV. 
}\label{fig:PT_disorder}
\end{figure}

Thus, to further characterize the emergence of many-body quantum chaos in vdW systems, we analyze statistical fluctuations of the excitonic oscillator  or dipolar strengths, $D$. This provides a complementary probe of ergodicity sensitive to eigenvector structure rather than spectral correlations alone.
Figs.~\ref{fig:PT_clean} and \ref{fig:PT_disorder} show representative distributions of the dipolar strengths associated with specific spectral regions. %
These distributions are compared to Eq. (\ref{Eq:PTnu}), with $\nu =1, 2$ (GOE, GUE limits) and also fitted by treating $\nu$ a free parameter.
For the pristine heterobilayer, Fig. \ref{fig:PT_clean}, %
the distribution of the dipolar strengths deviates substantially from both GOE and GUE distributions in each of the considered energy windows.
Instead, the best fit to  Eq. \eqref{Eq:PTnu} yields an effective number of degrees of freedom $\nu = 0.6$, with remaining deviations at the tails of the distribution (see SI). This value points towards anomalously strong fluctuations and incomplete randomization of the excitonic states, an observation also consistent with the eigenvalue analysis, where the ratios failed to develop a robust GOE plateau. In addition, this strengthens our conclusion that this excitonic Hamiltonian remains ``partially'' chaotic, due to the approximate valley symmetry that fragments the excitonic subspaces.
The lattice-disordered case, Fig. \ref{fig:PT_disorder}, exhibits a starkly different behavior. At lower energies, we find a good fit to Eq. \eqref{Eq:PTnu} with $\nu = 1.67$ indicating suppressed dipole fluctuations and more homogeneous intensities compared to the pristine case. Crucially, in the energy window where the spacing ratio approaches the GUE limit, the dipolar strength distribution follows closely the pure GUE Porter-Thomas law ($\nu = 2$). This demonstrates that at higher energies the excitonic states acquire fluctuations expected for chaotic wavefunctions and reflects the fact that the crossover but a genuine manifestation of wavefunction ergodicity in the many-body configuration space.

\begin{figure}
\centering
\includegraphics[width=0.95\columnwidth]{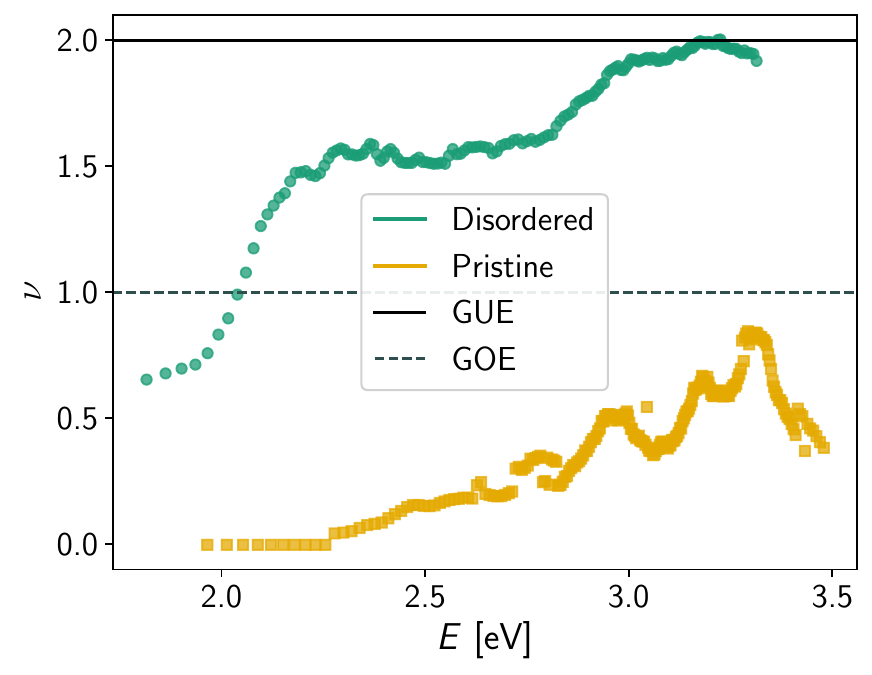}
\caption{Energy dependence of the Porter–Thomas fitting parameter $\nu$ for disordered (green circles) and pristine (yellow squares) cases. The black continuous line corresponds to $\nu = 2$ (GUE) and the dark grey dashed line to $\nu =1$ (GOE).
}\label{fig:nuE}
\end{figure}

Fig. \ref{fig:nuE} shows the energy dependence of the effective Porter-Thomas index, $\nu(E)$ across the entire excitonic spectrum. %
In the pristine system, $\nu(E) < 1$ (sub-GOE or incomplete ergodicity) persists throughout the entire excitonic bandwidth, even where the short range spectral correlations transiently approach the GOE limit. 
This indicates that the spectral signatures of level repulsion develop more rapidly than full wavefunction ergodicity. Such difference between eigenvalue and eigenvector statistics has previously been observed in nuclear, mesoscopic, and molecular systems \cite{Guhr1998, Alhassid2000, Molina2000, Gomez2011}.
In the contrast, for the disordered system the values of $\nu(E)$ are always larger and reach the $\nu(E) = 2$ value over a finite window, indicating that the oscillator strength fluctuations fully saturate the GUE limit. This behavior mirrors the emergence of the GUE plateau seen in Fig. \ref{fig:ratio_total} and confirms a gradual, energy-dependent transition from incomplete chaos in the pristine vdW system to a fully developed ergodic regime driven by disorder-enhanced intervalley mixing in the lattice-disorder case.

To complete our characterization of ergodicity, we investigate long-range spectral correlations \cite{Corps2021}, which are governed by the (excitonic) Thouless energy scale $E_{\textnormal{Th}} = \hbar / \tau_{\textnormal{Th}}$. Here, $\tau_{\textnormal{Th}}$ is the characteristic time required for a many-body excitation to diffusively explore the accessible excitonic configuration space.
This scale separates universal RMT correlations from system-specific spectral fluctuations.
We extract $E_{\textnormal{Th}}$  by evaluation the power spectrum of the cumulative level fluctuations $\delta_n$ (see End Matter and SI), which is particularly sensitive to long-range correlations and has been widely used  \cite{Relano2002,Relano2004,Faleiro2004,Santhanam2005,Gomez2011,Riser2017,Corps2020}.  %
\begin{figure}
\centering
\includegraphics[width=0.95\columnwidth]{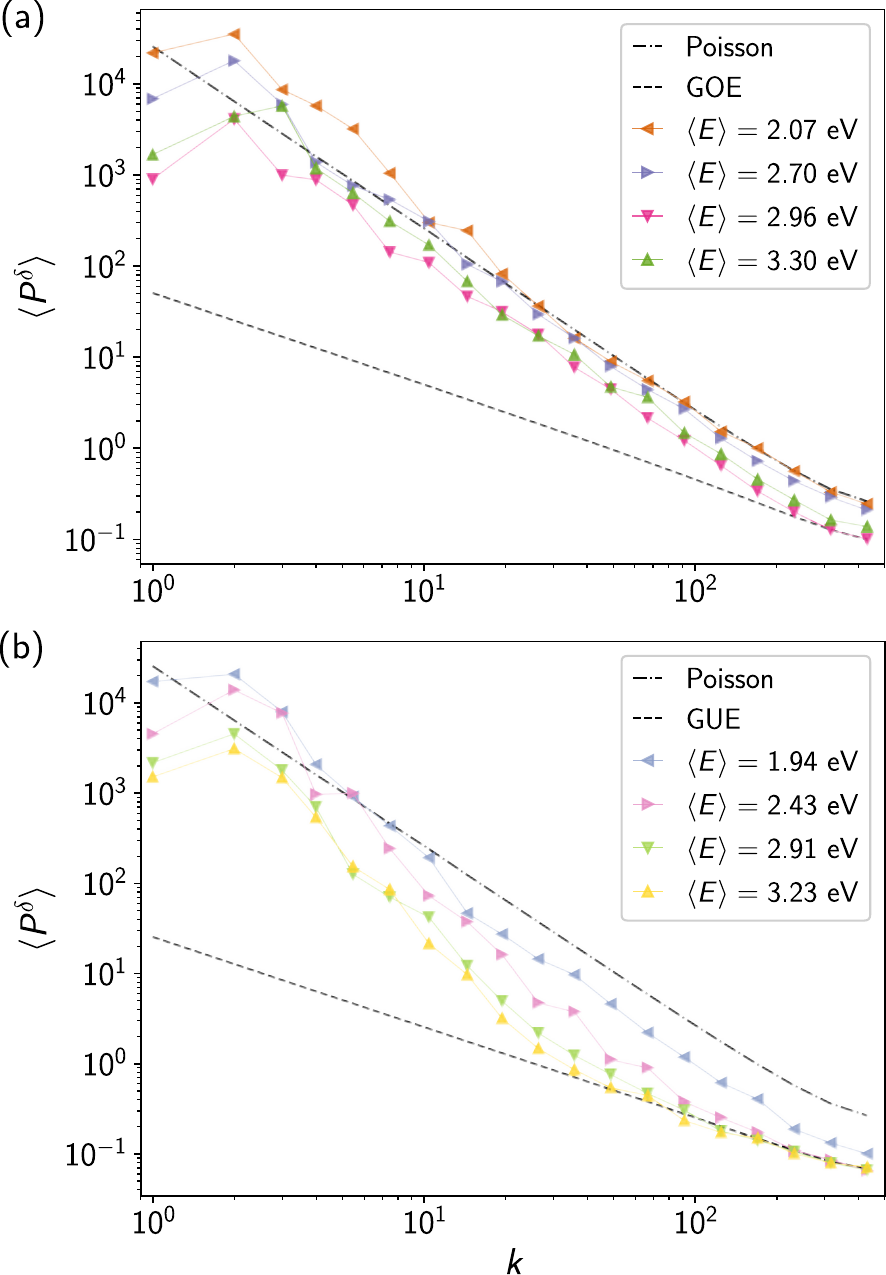}
\caption{Logarithmically averaged power spectrum of the cumulative fluctuation $\delta_n$ statistic for the (a) pristine  and the (b) lattice-disordered systems at different average energies $\langle E \rangle$. The dashed dotted line denote the Poisson limit and the dashed lines the RMT prediction for the GOE and GUE. 
The Thouless energy is extracted from the separation scale $k_{\mathrm{sep}}$, defined by the departure of the numerical data from the RMT prediction in the small-k (long-range) regime.}
\label{fig:power_spectrum}
\end{figure}
Fig. \ref{fig:power_spectrum} shows the logarithmically averaged power spectrum for pristine and disordered heterobilayers. The spectra are binned in $5000$ consecutive levels with five independently unfolded consecutive windows of $1000$ levels using a polynomial fit to density of states. The corresponding power spectra are then averaged over all windows and binned logarithmically in $k$ \cite{Corps2020, Corps2021}.
When short-range statistics approach the RMT regime, the power spectrum follows the corresponding RMT prediction up to a characteristic separation index, $k_{\textnormal{sep}}$, which defines the range of the universal spectral correlations \cite{Molina2010}. From this crossover, we estimate the Thouless scale $E_{\textnormal{Th}} \simeq N \Delta E/ k_{\textnormal{sep}}$, where $N$ is the number of levels in each window and $\Delta E$ is the mean level spacing. 

In the pristine case [panel (a)], the power spectrum approximately follows the Poisson prediction, showing a weak crossover to GOE only at large $k$.
In contrast, for the disordered heterobilayer [panel (b)], we find a transition from the Poisson curve to the GUE at lower $k$, which gives $E_{\textnormal{Th}} \sim 3 \cdot 10^{-4}$ eV for $\langle E \rangle=2.43$ eV and $E_{\textnormal{Th}} \sim 3 \cdot 10^{-5}$ eV for $\langle E \rangle=2.91$ eV and $\langle E \rangle=3.23$ eV. 
We note that although the separation scale $k_\textnormal{sep}$ increases at higher energies due to the larger density of states, the corresponding Thouless energy decreases as it scales with the rapidly shrinking mean level spacing. 
These values correspond to Thouless times  $\tau_{\textnormal{Th}} \simeq 0.22$ ps at $\langle E \rangle=2.43$ eV and $\tau_{\textnormal{Th}} \simeq 2.2$ ps in the deep chaotic regime. 
Characteristic times of the order $\simeq 0.1-1$ ps are remarkably fast and consistent with exchange, phonon, and disorder-assisted valley mixing \cite{Hao2016, Glazov2014}, being comparable to typical spectral mixing times in mesoscopic chaotic systems \cite{Beenakker1997}. This indicates that the dense, disordered excitonic spectrum undergoes fast ergodization across the entire many-body spectral bandwidth.

\subparagraph{\em Conclusions.} In this work, we have demonstrated that many-body exciton spectra
obtained from atomistic first-principles GW-BSE calculations in realistic 2D vdW interfaces
exhibit robust signatures of quantum chaos. 
We identified an energy-resolved crossover from regular to chaotic behavior  characterized by universal random-matrix statistics.
Our results reveal that the emergence and nature of the many-body quantum chaos is governed by symmetry. In the pristine vdW heterobilayer, spectral correlations evolve towards - but do not fully reach - the universal RMT limit consistent with a residual valley symmetry that constrains many-body level mixing.
In contrast, vacancies representing lattice-disorder induce in-gap flat bands that promote intervalley hybridization relaxing this constraint and driving the system towards fully ergodic regime at higher energies characterized by a different RMT universality class.
This behavior is consistently captured by the spacing ratio statistics, long-range correlations, and the energy dependence of the Porter-Thomas parameter extracted from the statistical fluctuations of the oscillator strength.
Importantly, the latter distributions provide a connection between quantum chaos and experimentally accessible optical observables. In the chaotic regime, many-body optical dipolar strength follow universal Porter–Thomas statistics, implying that excitonic quantum chaos should manifest as Ericson-type fluctuations in the optical absorption spectra \cite{Dirke1994}. 
Thus, this establishes a concrete route for observing quantum-chaotic behavior in interacting solid-state systems using standard optical spectroscopic techniques. More broadly, this work positions excitons and vdW matter as a realistic platform for exploring quantum chaos in interacting quantum matter and demonstrates that universal chaotic behavios can emerge
from first-principles many-body Hamiltonians. These findings open new opportunities to investigate the interplay between symmetry, disorder, hybridization, and ergodicity in low-dimensional quantum materials.

\begin{acknowledgments}
\textit{Acknowledgments.} D. H.-P. acknowledges insightful conversations with S. Bera.
D. H.-P. is grateful for funding from the Diputaci\'on Foral de Gipuzkoa through Grants 2023-FELL-000002-01, 2024-FELL-000009-01, 2025-FELL-000004-01 and 2025-CIE4-000036-01. 
D. H.-P. also acknowledges support of the Spanish {MICIU/AEI
/10.13039/501100011033} and FEDER, UE through Project No. PID2023-147324NA-I00 and from
IKUR Strategy, Quantum Technologies 2025 project \textit{M-Twist}. 
R. A. M. acknowledges support of the Spanish {MICIU/AEI
/10.13039/501100011033} and FEDER, UE through Project No. Grant PID2022-136285NB-C31 and support of the CSIC Research Platform on Quantum Technologies PTI-001.
\end{acknowledgments}
\vspace{0.2cm}
\textit{Data availability}. The data that support the findings of this article are not publicly available. The data are available from the authors upon reasonable request.

\section*{End Matter}
\subsection{\normalsize Mathematical tools}
\subparagraph{\em Excitons \& Bethe-Salpeter equation.}
The exciton spectra were obtained by numerically solving the Bethe-Salpether equation \cite{Salpeter1951, Rohlfing1998, Rohlfing2000} (see also Supporting Information and references therein)
\begin{equation}
    (E_{c\mathbf{k}} - E_{v\mathbf{k}}) A_{vc\mathbf{k}}^{(i)} + \sum_{v'c' \mathbf{k}'} K^{\textnormal{eh}}_{vc\mathbf{k}; v'c'\mathbf{k}'} A_{v'c'\mathbf{k}'}^{(i)}  =\varepsilon_i A_{vc\mathbf{k}}^{(i)}.\label{e3}
\end{equation}
Here, $E_{c\mathbf{k}}$ (resp. $E_{v\mathbf{k}}$) corresponds to the quasiparticle corrected conduction (resp. valence) bands, $K^{\textnormal{eh}}_{vc\mathbf{k}; v'c'\mathbf{k}'} $ are the matrix elements of the electron-hole interaction kernel - composed by an attractive screened direct and a repulsive bare exchange Coulomb terms, $\varepsilon_i$ is the exciton energy and $A^{(i)}_{vc\mathbf{k}}$  the amplitude of the exciton state. In real space, the exciton wavefunction takes the form
\begin{equation}
 \Psi^{(i)}(\mathbf{r}_e, \mathbf{r}_h) = \sum_{vc\mathbf{k}} A^{(i)}_{vc\mathbf{k}}  \psi^\ast_{v\mathbf{k}}(\mathbf{r}_h) \psi_{c\mathbf{k}}(\mathbf{r}_e),
\end{equation}
with $\psi_{c\mathbf{k}}(\mathbf{r}_e)$ being the spinor wavefunction describing for the electron at position $\mathbf{r}_e$ and quantum numbers $c,\mathbf{k}$ (band index and crystal momentum respectively). Correspondingly, $\psi_{v\mathbf{k}}(\mathbf{r}_h)$ is the spinor for the hole at position $\mathbf{r}_h$, valence band index $v$ and momentum $\mathbf{k}$.
The (many-body) oscillator or dipolar strengths, $D_i$, are computed using BerkeleyGW \cite{Deslippe2012} as
\begin{equation}
D_i = \dfrac{2|\mathbf{e} \cdot \langle 0 | \mathbf{v} | \Psi^{(i)}\rangle |^2}{\varepsilon_i},
\end{equation}
with $| 0 \rangle$ being the ground state, and $\langle 0 | \mathbf{v} | \Psi^{(i)}\rangle $ the transition matrix element of the velocity operator between the ground state (non-bounded electron-hole pairs) and the exciton state $i$.
%

\subparagraph{\em Consecutive spacing ratios.} The study of quantum chaos seeks to identify signatures of chaotic dynamics in quantum systems, including systems do not possess a classical counterpart.
A central tool for this purpose is spectral statistics, which probes correlations between the eigenvalues of quantum Hamiltonians and enables comparison with the universal predictions of RMT \cite{Bohigas1984,BerryTabor1977,Mehta2004,Haake_book}.
In complex systems, however, conventional level spacing analysis requires unfolding the spectrum, a nontrivial procedure that can introduce systematic artifacts \cite{Gomez2002}.
To circumvent this issue, we employ the consecutive spacing ratio \cite{Oganesyan2007,Atas2013}
\begin{equation}
    r_n = \dfrac{ \min(s_n, s_{n+1})}{\max(s_n, s_{n+1})},
\end{equation}
with $s_i=\varepsilon_{i+1}-\varepsilon_i$.
Spacing ratios preserve the universal spectral correlations predicted by RMT while being insensitive to the local density of states. Thus, they are particularly well suited for systems with \textit{irregular} spectra such as finite-size many-body atomistic Hamiltonians (this work), Floquet Hamiltonians or disordered Hamiltonian models.
%

\subparagraph{\em Porter-Thomas statistics.} In random-matrix theory, chaotic eigenstates behave as random vectors, leading to universal statistics of transition amplitudes and intensities described by Porter-Thomas distributions \cite{Porter1956} [see Eq. \eqref{Eq:PTnu}]. These generalized $\chi^2$ distributions were originally introduced to model transition-rate fluctuations in compound nuclei and since has become a standard tool for understanding quantum chaos and ergodicity in complex quantum systems \cite{Shriner1987,Camarda1983,Delon1991,Flambaum1998,Arute2019,Claeys2025}.
Consider a normalized excitonic state with expansion coefficients $A_\alpha$, where $\alpha$ correspond to the exciton quantum numbers. We represent the ensemble of wavefunction intensities by $x = \{N |A_\alpha|^2\}$, where $N$ is the dimension of the Hilbert-space.
For chaotic systems described by the GOE, the wavefunction intensities are real and the distribution takes the form
\begin{equation}\label{eq:PT_GOE}
P_1(x) = \frac{1}{\sqrt{2\pi x_0 x}} \exp\!\left(-\frac{x}{2x_0}\right),
\end{equation}
which is a $\chi^2$ distribution with $\nu = 1$ degrees of
freedom.
For systems corresponding to the GUE, the real and imaginary parts of amplitudes fluctuate independently. The intensity
distribution then becomes
\begin{equation}\label{eq:PT_GUE}
P_2(x) = \frac{1}{x_0}\exp\!\left(-\frac{x}{x_0}\right),
\end{equation}
corresponding to a $\chi^2$ distribution with $\nu =2$ degrees of freedom.
These two distributions arise from the central limit behavior of chaotic states, whose components can be regarded as weakly correlated random variables \cite{Berry1977b}. Consequently, agreement of wavefunction derived quantities, such as the oscillator strength with, Eqs. \eqref{eq:PT_GOE} and \eqref{eq:PT_GUE} yield a probe of universality of the eigenstates in the sense of RMT.

\subparagraph{\em Spectral long range correlations.}
We efficiently characterize the long-range spectral correlations \cite{Tekur2018, Rao2020} from the unfolded excitonic spectrum $\{\varepsilon_i\}_{i=1}^N$, normalized such that the mean level spacing $\langle s\rangle=1$. Defining the nearest-neighbor level spacings, $s_i=\varepsilon_{i+1}-\varepsilon_i$, the cumulative level fluctuation \cite{Corps2021}, $\delta_n$, is given by
\begin{equation}
\delta_n=\sum_{i=1}^{n}\bigl(s_i-1\bigr)=\varepsilon_{n+1}-\varepsilon_1-n,
\end{equation}
with $n=1,\dots,N-1$.
The discrete power spectrum of \(\delta_n\) is then defined as
\begin{equation}
S(k)\equiv \frac{1}{N}\left|\sum_{n=1}^{N-1}\delta_n\,e^{-2\pi i kn/N}\right|^2,
\end{equation}
for $k=1,\dots,\left\lfloor{N}/{2}\right\rfloor$.
This quantity probes correlations over a characteristic scale $\ell \sim N/k$: Large values of $k$ are sensitive to short-range correlations (small $\ell$), while small values of $k$ probe long-range correlations.
In practice, $S(k)$ is computed within independently unfolded spectral windows (typically containing $N=1000$ levels), the ensemble average calculated over multiple consecutive windows to reduce statistical fluctuations and, when appropriate, further averaged using logarithmic binning in $k$ to better highlight scaling regimes.

RMT provides parameter-free predictions for the ensemble averaged power spectrum, $\langle S(k)\rangle$, which can be expressed in terms of the spectral form factor $K(\tau)$, with $\tau=k/N$.
For the ensemble averaged power spectrum of the level counting fluctuations, RMT yields
\begin{align}
\bigl\langle P_k^{n}\bigr\rangle &=
\frac{N^{2}}{4\pi^{2}}\left[\frac{K(k/N)-1}{k^{2}} + \frac{K(1-k/N)-1}{(N-k)^{2}}\right]
\notag \\ & +\frac{1}{4\sin^{2}(\pi k/N)}. \label{e4}
\end{align}
Here, for Poisson statistics, \(K(\tau)=1\), for the GUE,
\begin{equation}
K(\tau)=
\begin{cases}
 \tau, & \tau \leq 1, \\
 1, & \tau \geq 1,
\end{cases}
\end{equation}
while for the GOE
\begin{equation}
K(\tau)=
\begin{cases}
2\tau-\tau\ln(1+2\tau), & \tau\le 1,\\[4pt]
2-\tau\ln\!\left(\dfrac{2\tau+1}{2\tau-1}\right), & \tau\ge 1.
\end{cases}
\end{equation}
Eq. \eqref{e4} allows for the calculation of the RMT ensemble averaged discrete power spectrum
\begin{equation}
\bigl\langle S(k)\bigr\rangle_\textnormal{RMT} =
\begin{cases}
\bigl\langle P_k^{n}\bigr\rangle, & \text{Poisson (integrable)},\\[4pt]
\bigl\langle P_k^{n}\bigr\rangle-\dfrac{1}{12}, & \text{GOE/GUE (chaotic)},
\end{cases}
\end{equation}
which is related to the power spectrum of the level-number fluctuations by a constant shift.
Deviations of \(\langle S(k)\rangle\) from the GOE/GUE prediction at low $k$ define a crossover index, $k_{\mathrm{sep}}$. This index can be used to estimate the Thouless scale in ``level'' number $\ell_{\mathrm{Th}}\sim N/k_{\mathrm{sep}}$ and, through the mean level spacing in original energy spectrum, $\Delta E$, the correspoding Thouless energy $E_{\mathrm{Th}}\sim \ell_{\mathrm{Th}}\Delta E$.

\bibliography{biblio}
\end{document}